\documentclass{webofc}
\usepackage[varg]{txfonts}   % Web of Conferences font
\usepackage{lineno}
\newcommand{\lam}{\mbox{$\Lambda$}\xspace}
\newcommand{\alam}{\mbox{$\bar{\Lambda}$}\xspace}

\newcommand{\sqsn}{\mbox{$\sqrt{s_{_{NN}}}$}\xspace}
\newcommand{\pt}{\mbox{$p_T$}\xspace}
\begin{document}
%\linenumbers
%
\title{Global polarization and spin alignment measurements}
%
% subtitle is optionnal
%%%\subtitle{Do you have a subtitle?\\ If so, write it here}

\author{\firstname{Takafumi} \lastname{Niida}\inst{1,3}\fnsep\thanks{\email{niida@bnl.gov}} 
}

\institute{University of Tsukuba, 1-1-1 Tennoudai, Tsukuba, Ibaraki 305-8571, JAPAN
%\and
}

\abstract{%
Recent experimental results on global polarization of hyperons and spin alignment of vector mesons in heavy-ion collisions are reviewed,
especially in context of the energy dependence and particle species dependence.
New results on local polarization along the beam direction at the LHC are discussed with previous measurements at RHIC.
Future outlook of these measurements and possible future directions for better understanding of local vorticity field are briefly discussed.
}
\maketitle
\section{Introduction}
\label{intro}
In non-central heavy-ion collisions, an initial orbital angular momentum carried by two colliding nuclei is partially converted 
to the spin angular momentum of particles produced in the collisions~\cite{Liang:2004ph,Voloshin:2004ha,Becattini:2007sr}. This phenomenon is called global polarization, 
i.e. produced particles are globally polarized along the direction of the initial orbital angular momentum which is perpendicular to reaction plane.
The observation of \lam global polarization by the STAR Collaboration~\cite{STAR:2017ckg}, indicating the creation of vortical fluid, opens new direction to study a nuclear matter 
not only under high temperature and/or baryon density but also under the fastest vorticity field~\cite{Jiang:2016wvv,Fujimoto:2021xix} 
as well as possibly to study the initial strong magnetic field ($B\sim10^{13}$~T) and its lifetime which has large uncertainty~\cite{McLerran:2013hla}.

While results on the average global polarization are well reproduced by theoretical models~\cite{Karpenko:2016jyx,Li:2017slc,Sun:2017xhx,Xie:2017upb,Ivanov:2019ern} 
in a wide range of collision energies, differential measurements and local polarization measurements~\cite{STAR:2019erd,Niida:2018hfw,Becattini:2020ngo} found discrepancies with models which still remain open questions.
In this proceedings, recent experimental updates on global and local polarization measurements of hyperons as well as spin alignment measurements of vector mesons 
are reviewed and their physics implications are discussed.

\section{Hyperon global polarization}\label{sec-1}
\subsection{\lam global polarization}
\begin{figure}[h]\vspace{-0.1cm}
\centering
\includegraphics[width=8cm,clip]{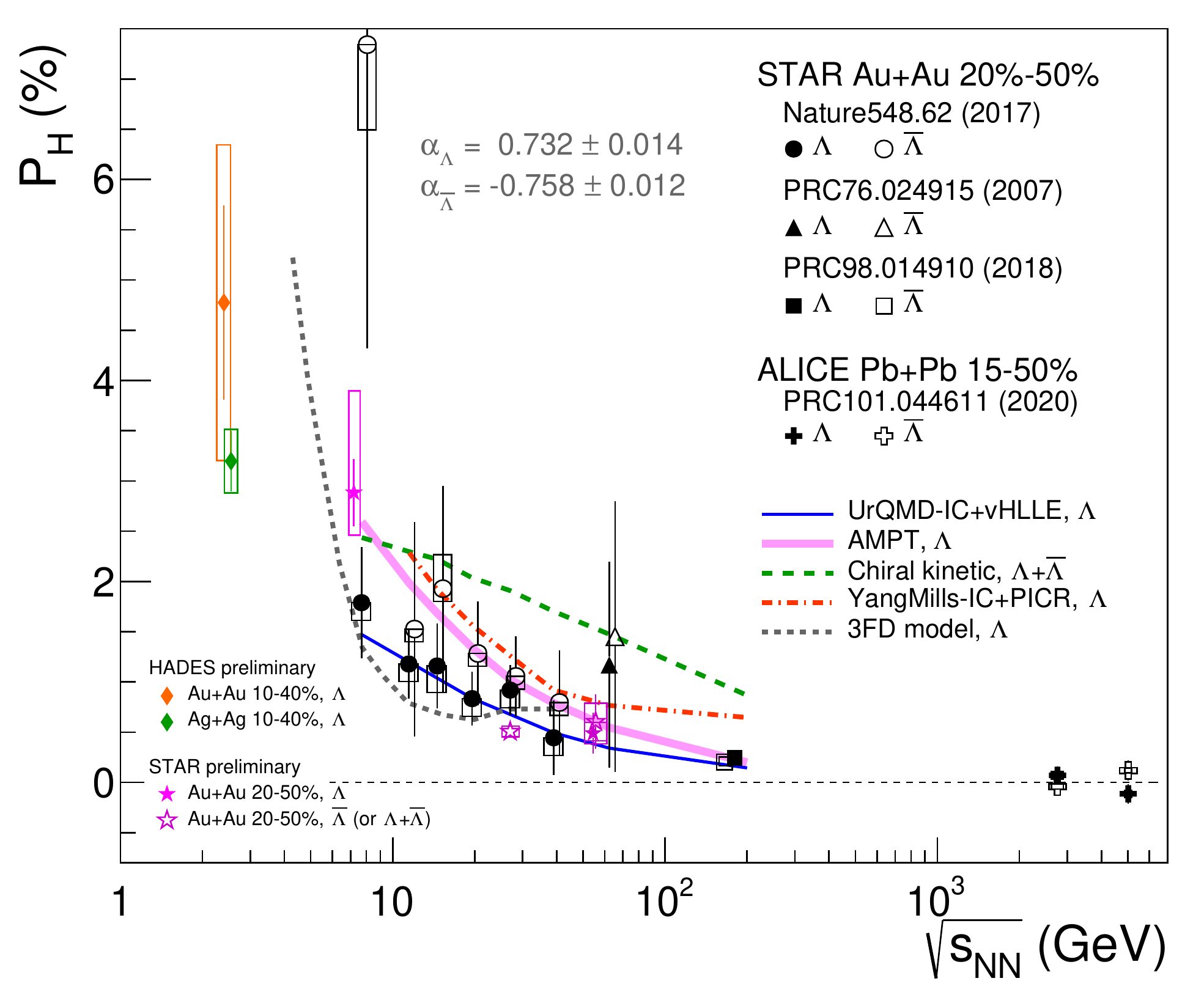}\vspace{-0.1cm}
\caption{Energy dependence of global polarization of \lam hyperons in heavy-ion collisions, 
comparing with various theoretical calculations~\cite{Karpenko:2016jyx,Li:2017slc,Sun:2017xhx,Xie:2017upb,Ivanov:2019ern}.
The experimental data are taken from Refs.~\cite{STAR:2017ckg,STAR:2018gyt,sqm2021:Okubo,sqm2021:Kornas,ALICE:2019onw} and 
rescaled with new decay parameter~\cite{ParticleDataGroup:2020ssz} shown in the figure.}
\label{fig:PHvsRootS}   % Give a unique label
\end{figure}

Polarization can be studied using hyperons' weak decay, by analyzing angular distribution of decay product: $dN/d\cos\theta^{\ast}\propto 1+\alpha_{H}P_H\cos\theta^\ast$
where $\theta$ denotes a polar angle of daughter baryon relative to the direction of hyperon polarization $P_H$, 
$\alpha$ is decay parameter ($\alpha_{\Lambda}\approx0.732$~\cite{ParticleDataGroup:2020ssz}), and the asterisk indicates analysis in the hyperon rest frame.
Global polarization of $\Lambda$ hyperons was first observed by STAR Collaboration in the beam energy scan of Au+Au collisions at \sqsn = 7.7--39 GeV~\cite{STAR:2017ckg}, 
and later confirmed at 200 GeV with better precision~\cite{STAR:2018gyt}. 
%The results show an energy dependence, increasing toward lower energies, 
%which could be understood by rapidity dependence of the initial shear flow and baryon stopping at lower energies with finite detector acceptance.
%Longer lifetime of the system at higher energy may also dilute the polarization, leading to the similar energy dependence.
%Many theoretical models such as hydordinamic and tranport models can explain well the measured polarization averaged over phase-space, 
%while there exists open questions especially in differential measurements, e.g. descrepancy in azimuthal angle dependence between the data and models and unknown rapidity dependence.
%
%Figure~\ref{fig:PHvsRootS} shows a compilation of results on \lam (\alam) global polarization comparing with various theoretical calculations, 
%including new preliminary results from the STAR and HADES Collaborations~\cite{sqm2021:Kornas}.
%Based on theoretical and phenominolgical expectations, the polarization at the LHC energy would be of the order of 0.1\%. 
%Results in Pb+Pb collisision at \sqsn = 2.76 and 5.02 TeV from ALICE~\cite{ALICE:2019onw} are consitent with zero and the current uncertainty is still in the level of the expected sinal,
%which can be explored with large statistics in LHC-Run3.
%
As shown in Fig.~\ref{fig:PHvsRootS}, the results show an energy dependence, increasing toward lower energies, 
which could be understood by rapidity dependence of the initial shear flow field and baryon stopping at lower energies with finite detector acceptance.
Longer lifetime of the system at higher energy may also dilute the polarization, leading to the similar energy dependence.
Various theoretical models~\cite{Karpenko:2016jyx,Li:2017slc,Sun:2017xhx,Xie:2017upb,Ivanov:2019ern} such as hydrodynamic and transport models 
can explain well the measured polarization averaged over phase-space as shown in Fig.~\ref{fig:PHvsRootS}, 
while there exists open questions in differential measurements as mentioned, e.g. discrepancy in azimuthal angle dependence between the data and models and unknown rapidity dependence.
Based on phenomenological and theoretical expectations~\cite{Voloshin:2017kqp,Karpenko:2018erl}, the polarization at the LHC energy would be of the order of 0.1\%. 
Results in Pb+Pb collisions at \sqsn = 2.76 and 5.02 TeV from ALICE~\cite{ALICE:2019onw} are consistent with zero and the current statistical uncertainty is still in the level of the expected signal,
which can be explored with large statistics in LHC-Run3.

The polarization at lower energy is also of particular interest since the system would turn to hadronic matter from partonic matter when decreasing the energy. 
How does the initial orbital angular momentum couple with spins of partons or hadrons and any difference there? What is the relaxation time for spin-orbit coupling compared to the system lifetime?
New preliminary result at 7.2 GeV from STAR fixed-target program~\cite{sqm2021:Okubo} is found to follow the global trend as shown in Fig.~\ref{fig:PHvsRootS}. 
Furthermore, new preliminary results in Ag+Ag 2.55 GeV and Au+Au 2.4 GeV collisions 
from HADES experiment indicate comparable or even larger polarization compared to that at 7.2 GeV. A transport model (UrQMD)~\cite{Deng:2020ygd} predicts that the kinetic vorticity 
becomes maximum around 3 GeV and then decreases when approaching \sqsn = $2m_N$ where $m_N$ is the mass of nucleon. 
It is worth to mention that after SQM2021 conference the STAR Collaboration has reported new result on \lam global polarization at 3 GeV~\cite{STAR:2021beb}, 
which is comparable to or possibly larger than the HADES preliminary results.
Note that different detector acceptance lead to difference in the measured polarization if there is rapidity and/or $p_T$ dependence 
which seems to be not significant with current uncertainties.

\subsection{$\Xi$ and $\Omega$ global polarization}
%Global polarization measurement has been performed only for \lam and \alam hyperons so far. 
As mentioned in previous section, there still exists open questions and discrepancy between the data and models. Further experimental inputs, 
especially polarization measurements with other particles (e.g. different spin), will help to better understand the nature of 
the polarization in heavy-ion collisions. With the assumption of local thermal equilibrium of the system, the polarization of 
these hyperons $P$ is determined by the local thermal vorticity $\omega$ as $P=(s+1)\omega/T$ where $s$ is the spin of the particle and $T$ is the temperature.

\begin{figure}[h]\vspace{-0.1cm}
\centering
\includegraphics[width=8cm,clip]{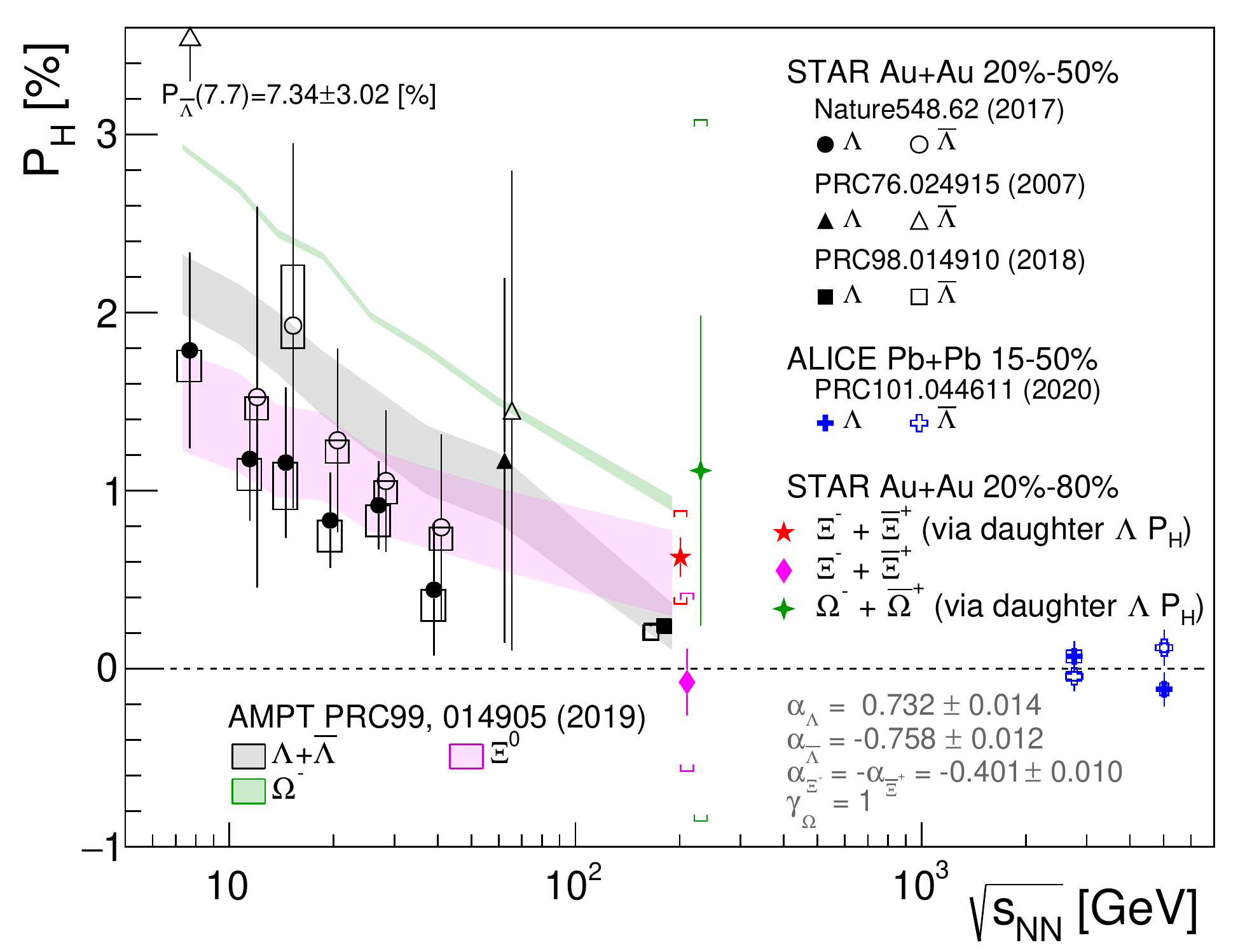}\vspace{-0.1cm}
\caption{$\Xi$ and $\Omega$ global polarization at 200 GeV with previous results for \lam (\alam) as a function of collision energy.
Calculations from AMPT model~\cite{Wei:2018zfb} are also shown with shaded bands. 
Previous results for \lam (\alam) polarization are rescaled with new decay parameter~\cite{ParticleDataGroup:2020ssz} shown in the figure.}
\label{fig:PHxi}       % Give a unique label
\end{figure}

The STAR Collaboration has recently extended the measurement to $\Xi$ (spin-1/2) and $\Omega$ (spin-3/2) hyperons~\cite{STAR:2020xbm}. 
One should note that the decay parameter is different for each hyperons: $\alpha_{\Xi}$=-0.401 and $\alpha_{\Omega}=0.0157$~\cite{ParticleDataGroup:2020ssz}. 
Smaller magnitude of the decay parameter makes the measurement difficult, especially for case of $\Omega$ it is practically impossible to measure $\Omega$ polarization in this way.
Instead one can utilize the following relation to measure the polarization of $\Xi$ and $\Omega$ hyperons:
${\bf P}^{\ast}_{P} = C_{PD}{\bf P}^{\ast}_{D}$
where ${\bf P}_{P(D)}$ denotes the polarization of parent (P) and its daughter (D) particles in the parent rest frame and 
$C_{PD}$ is the polarization transfer factor in the decay from the parent to the daughter. 
Results on $\Xi$ and $\Omega$ global polarization (particles and antiparticles combined) at 200 GeV from STAR~\cite{STAR:2020xbm} are shown in Fig.~\ref{fig:PHxi}, 
together with previous results for \lam and \alam. The $\Xi$ polarization was measured with the two independent methods: 
by analysis of daughter \lam distribution in $\Xi$ rest frame and by measuring polarization of daughter \lam taking into account the polarization transfer of $C_{\Xi\Lambda}=+0.944$.
Both results are consistent within uncertainties and the combined result show a slightly larger polarization than inclusive \lam, although the difference is not significant with the current uncertainties.
\vspace{-0.4cm}
Similarly, $\Omega$ polarization was measured via daughter polarization but there is an uncertainty in $C_{\Omega\Lambda}=(1+4\gamma_{\Omega})/5$ 
due to the unmeasured decay parameter $\gamma_{\Omega}$ which is likely either +1 or -1. The result based on $\gamma_{\Omega}=+1$ is shown in Fig.~\ref{fig:PHxi}.
The new results on $\Xi$ and $\Omega$ polarization, consistent with theoretical expectations as shown in the figure, reinforce the fluid vorticity and global polarization picture in heavy-ion collisions.
More precise measurements are anticipated at RHIC and the LHC, especially the uncertainty of the $\gamma_{\Omega}$ could be clarified assuming the vorticity picture 
and furthermore the measurement of $\Omega^{-}$ and $\bar{\Omega}^{+}$ separately would be interesting to look for the possible difference 
due to the initial magnetic field since $\Omega$ has a factor of three larger magnetic moment compared to \lam and $\Xi$.

%==========================================
\section{Spin alignment of vector mesons}
Vector mesons (spin-1) can be also used to study the particle polarization. Unlike hyperons, they decay via strong interaction where the parity is conserved, 
therefore one cannot know the direction of polarization.
The spin state of a vector meson can be described by $3\times3$ spin density matrix where there exists only one independent diagonal element $\rho_{00}$.
The diagonal element $\rho_{00}$ represents the probability to have spin projection onto a quantization axis to be zero. It can be studied by
measuring angular distributions of decay products of vector mesons: $dN/d\cos\theta^\ast\propto[1-\rho_{00}+(3\rho_{00}-1)\cos^2\theta^\ast]$ 
where $\theta^{\ast}$ is angle of decayed daughter relative to the polarization direction in the decay rest frame.
If the particles are unpolarized, $\rho_{00}=1/3$ on average. The deviation from $1/3$ indicates spin alignment of vector mesons.

\begin{figure}[h]\vspace{-1.0cm}
  \begin{minipage}{0.49\hsize}
  \centering
  \includegraphics[width=\linewidth,clip]{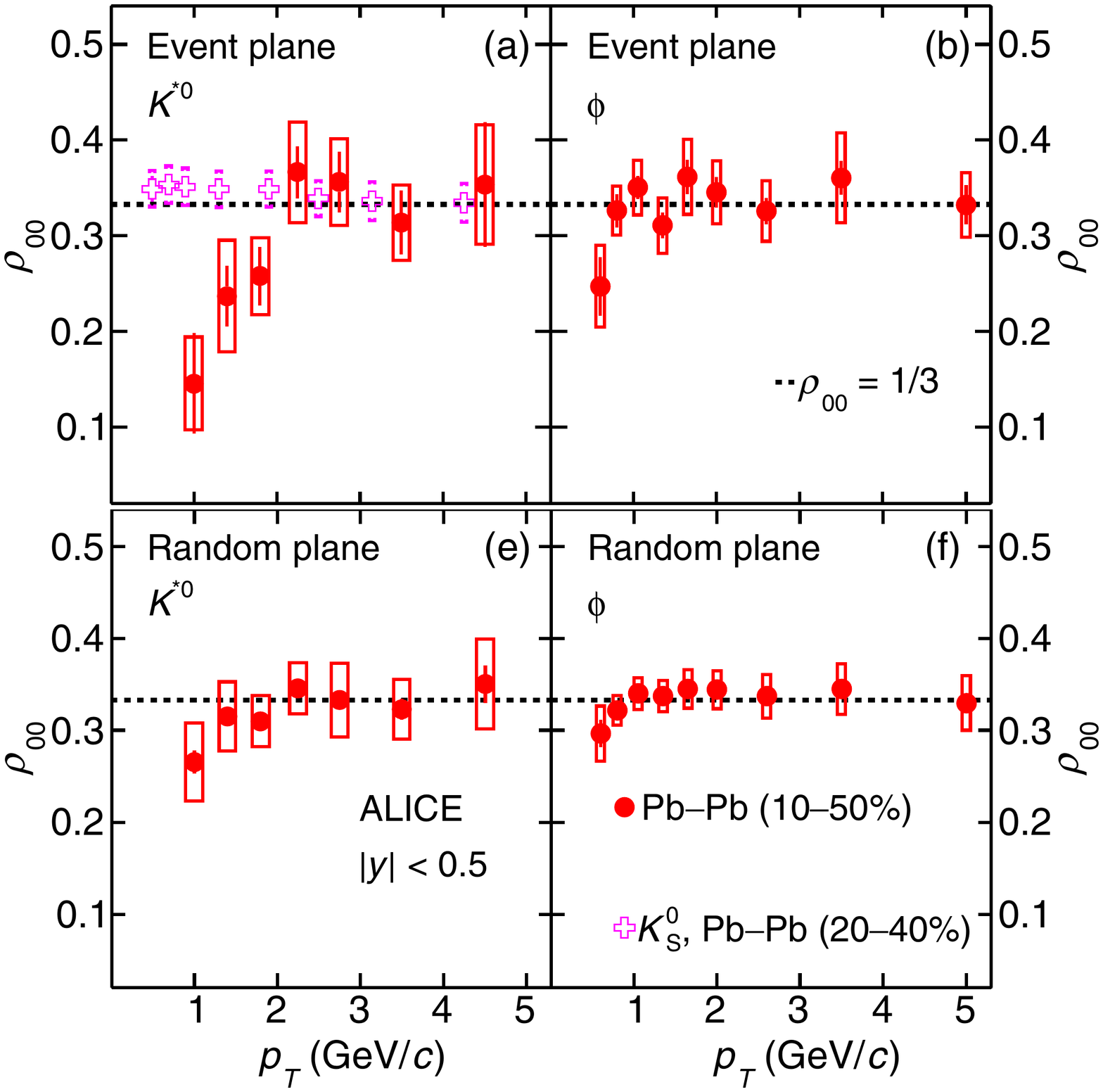}
  \end{minipage}
  \begin{minipage}{0.48\hsize}
  \centering
  \includegraphics[width=0.75\linewidth,clip]{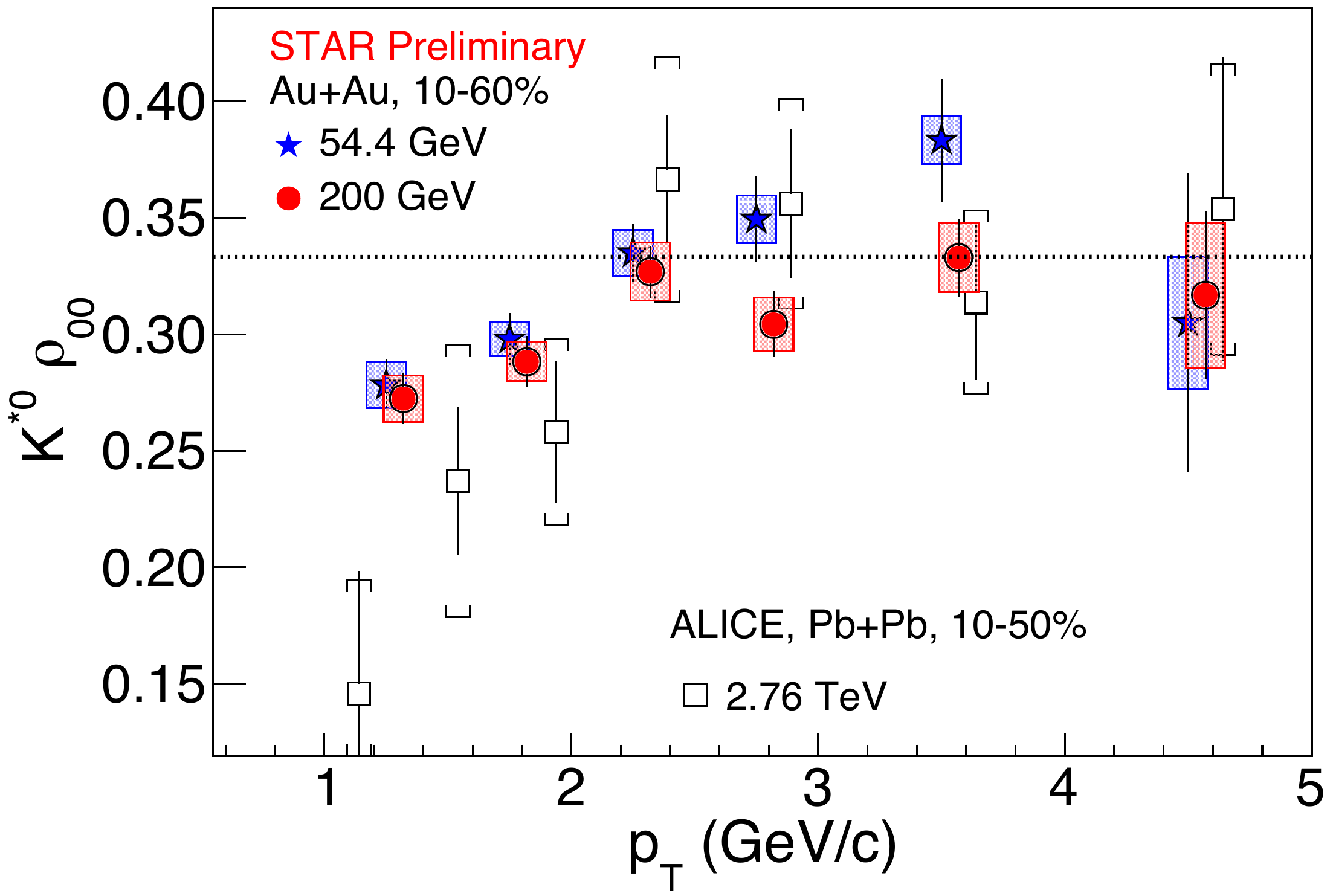}
  \includegraphics[width=0.75\linewidth,clip]{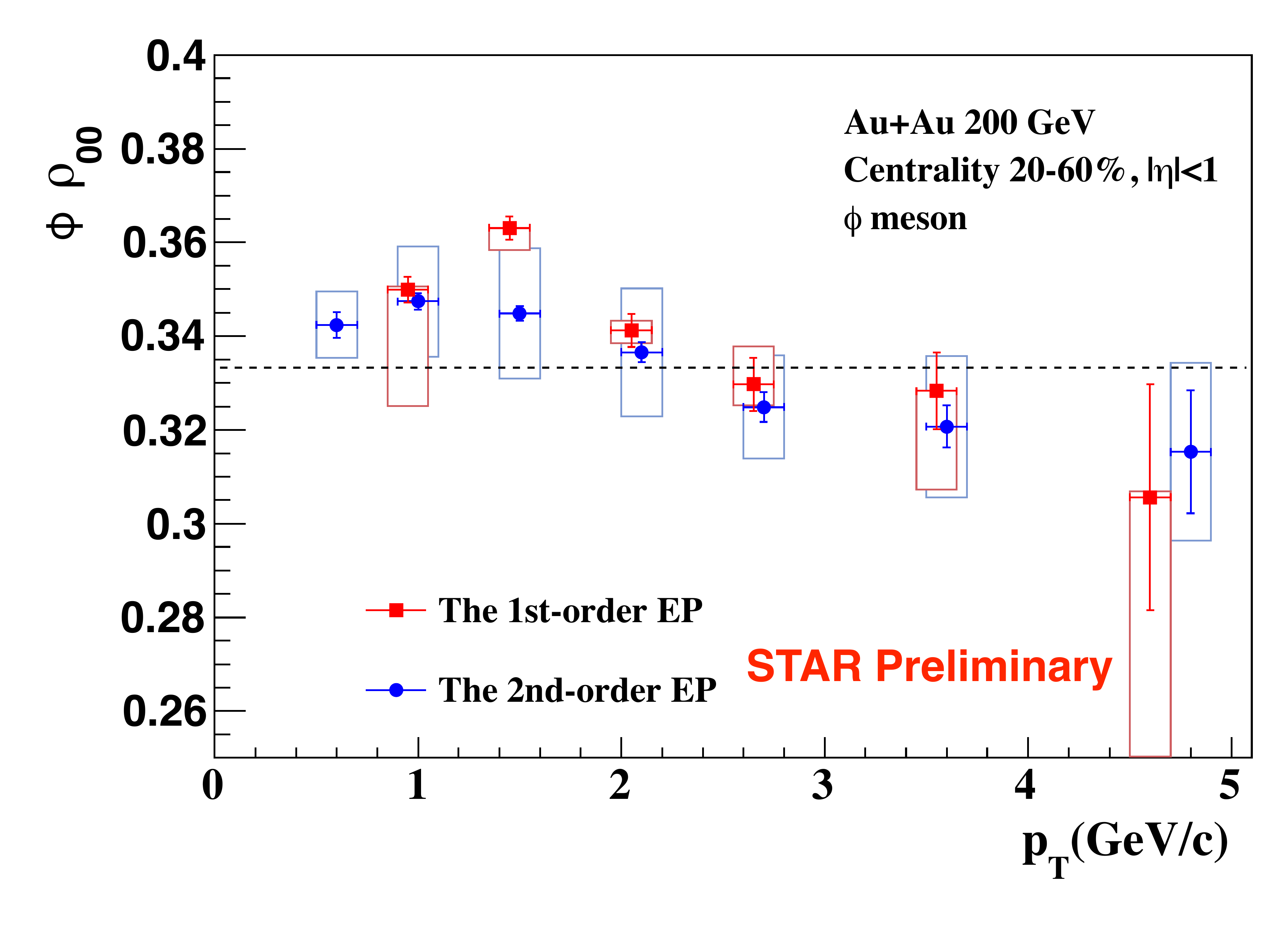}
  \end{minipage}\vspace{-1.2cm}
\caption{Average spin density matrix element $\rho_{00}$ of $K^{\ast0}$ and $\phi$ mesons in \sqsn = 2.76 TeV Pb+Pb collisions at the LHC (left)~\cite{ALICE:2019aid} 
	and in \sqsn = 54.4 and 200 GeV Au+Au collisions at RHIC (right)~\cite{Singha:2020qns,Zhou:2019lun}.}
\label{fig:rho00}       % Give a unique label
\end{figure}

Figure~\ref{fig:rho00} left panels show $\rho_{00}$ of $K^{\ast0}$ and $\phi$ mesons from ALICE experiment~\cite{ALICE:2019aid}.
The results relative to the event plane show a negative deviation from $1/3$ at $p_T<2$ GeV$c$, 
while the results of $K_{s}^0$ mesons (spin-0) are consistent with $1/3$ as expected.
Also, the results relative to random event plane serving as a baseline confirm the significance of the deviation.
Preliminary results from STAR experiment show similar behavior for $K^{\ast0}~\rho_{00}$ at \sqsn = 54.4 and 200 GeV~\cite{Singha:2020qns} as shown in the top-right panel of Fig.~\ref{fig:rho00}, 
while $\rho_{00}$ of $\phi$ mesons~\cite{Zhou:2019lun} show positive values at low $p_T$ as opposed to the negative deviation at the LHC.
Note that the contribution to the spin alignment from the vorticity is a second-order effect, i.e. $\rho_{00}\sim 1/[1+(\omega/T)^2]$. 
Therefore the observed deviation of $\rho_{00}$ cannot be explained by the vorticity picture based on the results on the \lam global polarization.
Also there seems no significant energy dependence in $K^{\ast0}~\rho_{00}$ unlike the \lam global polarization.
Such a large deviation should also contribute to the elliptic flow~\cite{Voloshin:2004ha}. Mean field of $\phi$ mesons may play a role but likely not for $K^{\ast0}~\rho_{00}$~\cite{Sheng:2019kmk}.
Further investigation is needed for better understanding the spin alignment measurements.
\vspace{-1cm}

\section{Local polarization}
The vorticity field may be more complicated varying with phase-space due to density fluctuations, energy deposit from jets, and collective flow.  
Theoretical calculations predict that the vorticity along the beam direction might be created due to the azimuthal anisotropic flow~\cite{Becattini:2017gcx,Voloshin:2017kqp}.
The STAR experiment observed the polarization of \lam (\alam) along the beam direction ($P_z$) in Au+Au collisions at \sqsn = 200 GeV, 
where the sign of $P_z$ changes with azimuthal angle as expected from the elliptic flow in origin.
However there is discrepancy in the sign between the data and models, even among various model calculations~\cite{Becattini:2020ngo}. 

\begin{figure}[h]
\centering
\includegraphics[width=0.48\linewidth,clip]{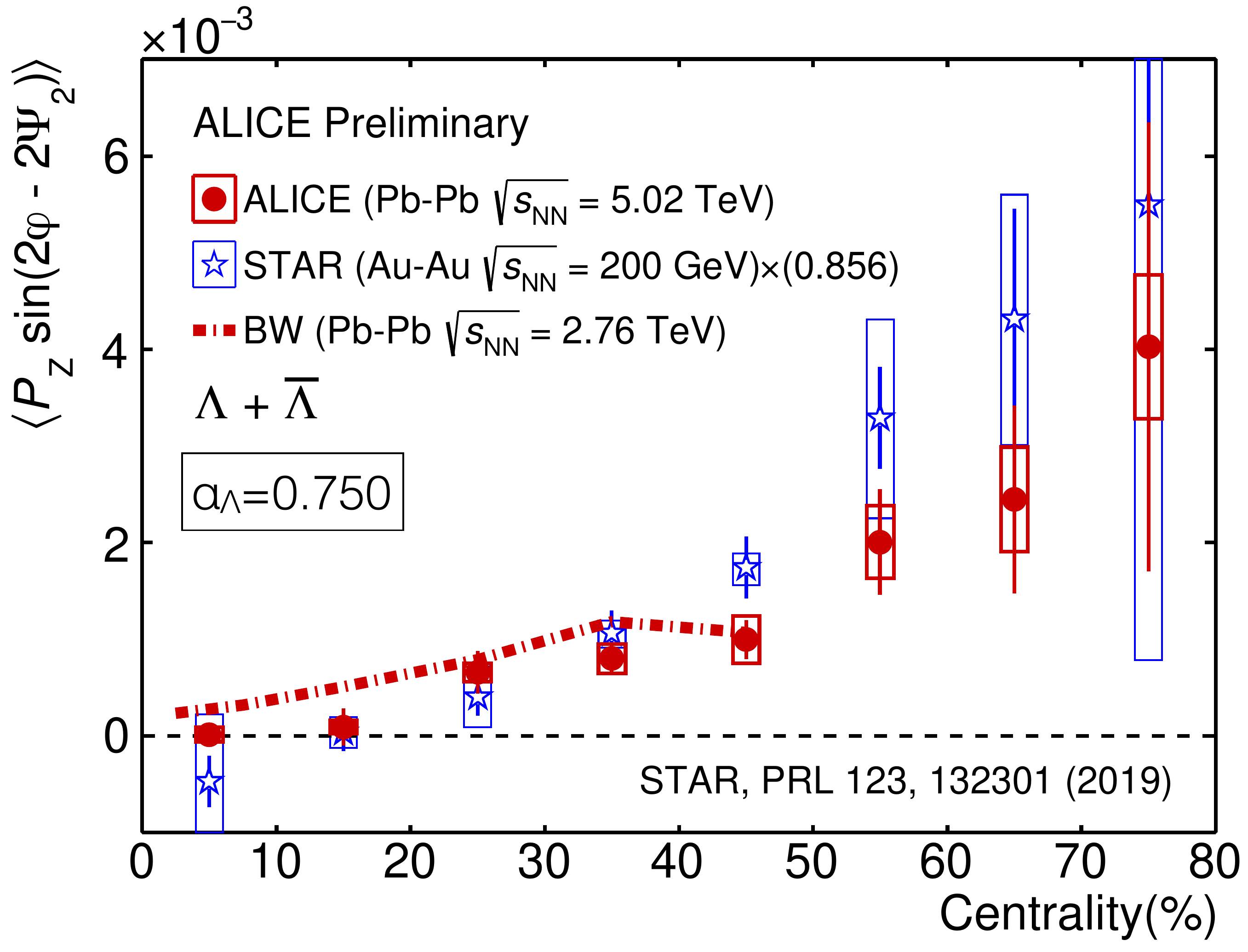}
\includegraphics[width=0.47\linewidth,clip]{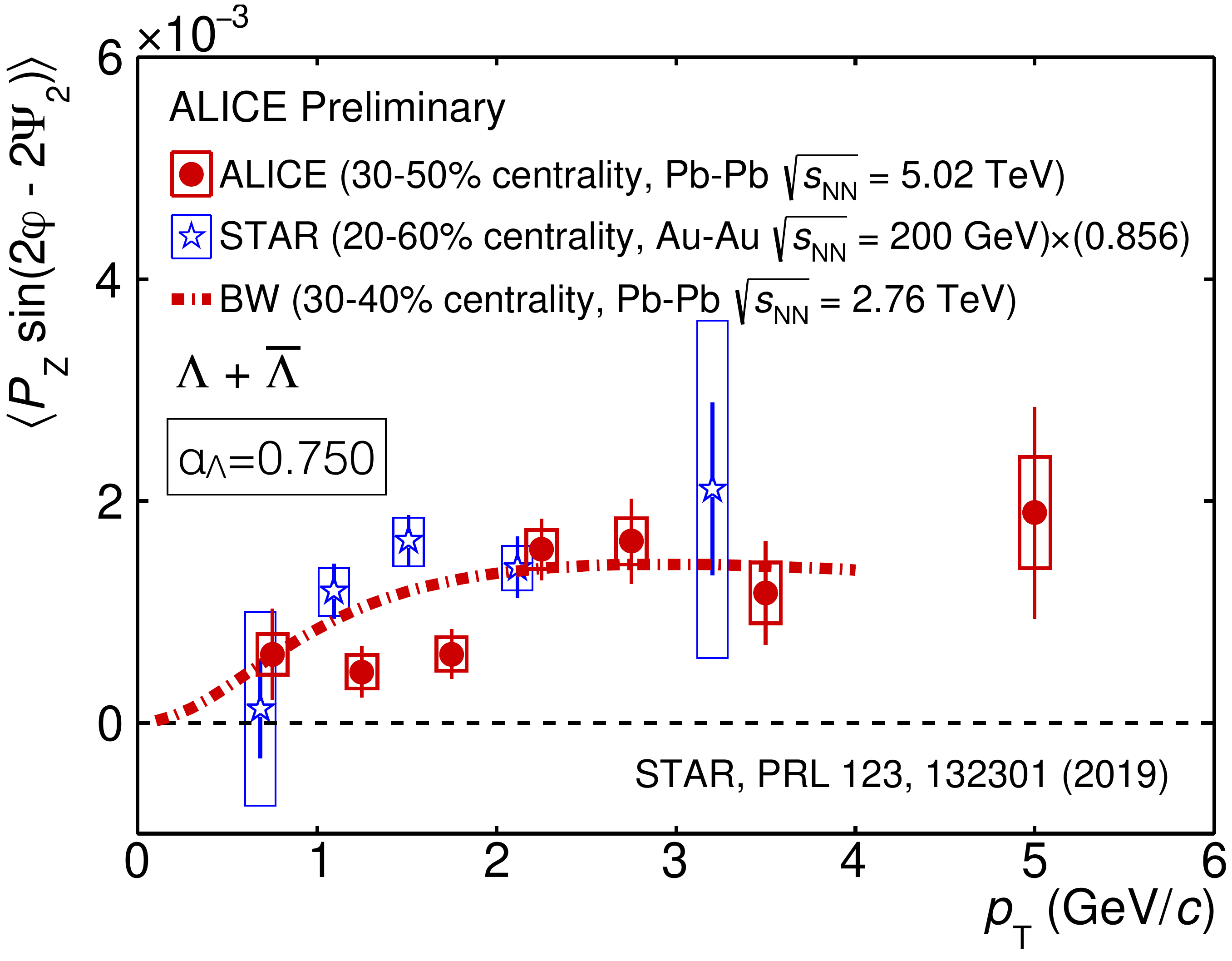}
\caption{Fourier sine coefficient of \lam+\alam polarization along the beam direction as a function of centrality (left) and transverse momentum \pt (right)
in Pb+Pb collisions at \sqsn = 5.02 TeV~\cite{sqm2021:Sarker} compared to those in Au+Au collisions at \sqsn = 200 GeV~\cite{STAR:2019erd}.}
\label{fig:Pz} 
\end{figure}

In this conference, new results in Pb+Pb collisions at \sqsn = 5.02 TeV~\cite{sqm2021:Sarker} were reported from the ALICE Collaboration as shown in Fig.~\ref{fig:Pz}.
\vspace{-0.4cm}
The results show similar behaviors and are comparable to the results from STAR; the sine modulation of $P_z$ increases towards peripheral collisions (Fig.~\ref{fig:Pz} left)
and shows a hint of $p_T$ dependence at $p_T<2$ GeV/$c$ (Fig.~\ref{fig:Pz} right). The blast-wave model calculations reproduce the data well in their sign and magnitude 
unlike more realistic hydrodynamic models. It should be noted that recent theoretical developments that consider contribution from shear tensor 
to the polarization in addition to thermal vorticity may explain the experimental data~\cite{Becattini:2021iol,Fu:2021pok}. 

Similar to the elliptic flow, one can naively expect that the longitudinal polarization may arise from higher-order flow, which can be explored with high statistics data in the future data taking.
Furthermore troidal structure of vorticity field is expected in central asymmetric collisions such as Cu+Au, d+Au, and p+Au collisions~\cite{Voloshin:2017kqp,Lisa:2021zkj}.
In this case, the direction of the polarization is along the azimuthal angle $\vec{\phi}$ of a hyperon. 
%$\vec{p_T}\times\vec{z}$ where $\vec{p_T}$ is momentum of a hyperon and $\vec{z}$ is the beam direction.
Future data with high statistics will allow us to study this phenomena.

\section{Summary}
This paper has presented recent experimental results on hyperon polarization and spin alignment of vector mesons.
New results on global polarization of \lam hyperons indicate that the polarization still increases towards \sqsn = 2.4-7 GeV. 
With given large uncertainties, it is still not clear whether the polarization starts to decrease as expected from theoretical calculations.
The first measurements on $\Xi$ and $\Omega$ global polarization reinforce the vorticity picture in heavy-ion collisions. 
To confirm the possible mass and/or spin dependence of the polarization, more precise measurements are needed.
Polarization of \lam hyperons along the beam direction as well as significant deviation of $\rho_{00}$ from $1/3$ were observed for $K^{\ast0}$ and $\phi$ mesons 
both at RHIC and the LHC but there still exists open questions.
Future runs at RHIC and the LHC as well as the data of the beam energy scan program phase-II from STAR will hopefully clarify those questions together with theoretical developments. 

%For two-column wide figures use syntax of figure~\ref{fig-2}
%\begin{figure*}
%\centering
%% Use the relevant command for your figure-insertion program
%% to insert the figure file. See example above.
%% If not, use
%\vspace*{5cm}       % Give the correct figure height in cm
%\caption{Please write your figure caption here}
%\label{fig-2}       % Give a unique label
%\end{figure*}
%
%For figure with sidecaption legend use syntax of figure
%\begin{figure}
%% Use the relevant command for your figure-insertion program
%% to insert the figure file.
%\centering
%\sidecaption
%%\includegraphics[width=5cm,clip]{tiger}
%\caption{Please write your figure caption here}
%\label{fig-3}       % Give a unique label
%\end{figure}

The author thanks the organizers of Strangeness in Quark Matter 2021 for the invitation to give this plenary talk. 
The author thanks S. Voloshin, S. Esumi, M. Lisa, F. Becattini, Y. Karpenko, and X.-G. Huang for fruitful discussions.

% BibTeX or Biber users please use (the style is already called in the class, ensure that the "woc.bst" style is in your local directory)
\bibliography{ref_sqm2021}

% Non-BibTeX users please use
%\begin{thebibliography}{}
%\bibitem{RefJ}
%% Format for Journal Reference
%Journal Author, Journal \textbf{Volume}, page numbers (year)
%% Format for books
%\bibitem{RefB}
%Book Author, \textit{Book title} (Publisher, place, year) page numbers
%\end{thebibliography}

\end{document}